# Acceleration of particles in an isotropic random force field

Hector Javier Durand-Manterola
Departamento de Ciencias Espaciales
Instituto de Geofisica
Universidad Nacional Autonoma de Mexico

hdurand_manterola@yahoo.com

Key Words: Physics, Particle Stochastic Acceleration.

**Abstract**

If we have a particle immersed in a field of random forces, each interaction of the particle with the field can enlarge or diminish its kinetic energy. In this work is shown that in general, for any field of random force with uniform distribution of directions, the probability to gain kinetic energy is larger that the probability to lose it. Therefore, if the particle is submitted to a great number of interactions with the force stochastic field, the final result will be that the particle will gain energy. The probability to gain energy in each interaction is

$$P_g = \frac{1}{2}\left(1 + \frac{T}{2P_0}\right)$$

Where T is the impulse given by the field and $P_0$ is the momentum of the particle before the interaction.
  The probability to lose energy in each interaction is:

$$P_l = \frac{1}{2}\left(1 - \frac{T}{2P_0}\right)$$

# 1 Introduction

A particle can be accelerated for many mechanisms; waves of plasma, shock waves, electric fields, and gravitational fields. All these mechanisms can be grouped in two classes: deterministic and stochastic (Jokipii, 1979).

Fermi (1949) suggested a mechanism to explain the high energies and the origin of the cosmic rays. He proposed that the charged particles of the interstellar environment are reflected by mirror-type magnetic configurations of the galactic magnetic field. And gain or lose energy depending if magnetic mirror is bringing near or moving away.

Fermi argued that, since it is easier that a particle collide with something that comes to their encounter that with something that moves away of it, then the probability of forward collision is larger that the probability of collision from behind and therefore the particles on the average will be accelerated. This random process is called Fermi acceleration of second order; due to that the average energy gained by interaction depends on the square of the velocity of the mirror. With this mechanism, Fermi was the first one that realized that, with a stochastic phenomenon, it could accelerate particles to very high energies. Subsequently several improvements were made to his idea. At present convincing models exist that describe stochastic acceleration in various environments (He, 2003; Bergstrom y Goobar 2004, pp 222). The stochastic acceleration is a mechanism that can accelerate particles in intergalactic and interstellar environment, in the Sun, in the interplanetary medium and in a planetary magnetosphere (Perez-Enriquez and Durand, 1992; Pérez-Peraza et al. 1994; Barbosa, 1994; Gallegos-Cruz et al. 1995; Martínez-Gómez et al. 2006, 2007a, 2007b). Along more than 50 years have been published hundreds of articles and books on the theme. Sagdeev et al. (1988) distinguish between stochastic acceleration, and stochastic heating. When the distribution of particles is maintained as before, but is widened, this is stochastic heating. But when the entire distribution is shifted toward higher speeds, we have stochastic acceleration, which can go accompanied also of heating

If we have a particle immersed in a field of random forces, each interaction with the field can enlarge or diminish its kinetic energy. But the diverse models that have been studied through the years, since that of Fermi in 1949 to the present models (v.g., Martínez-Gómez et al. 2006, 2007a, 2007b); each have a special distribution of stochastic field to accelerate particles. It could be thought that this acceleration is due to the particular configuration of forces or fields in each case, and with other configurations there will not be gain, and even could have loss.

In this work is demonstrated analytically, for the general case (i.e with any configuration of stochastic field) the very important fact that, if a particle is subject to the effects of a field of random force, the probability of gain energy is greater that the probability of loss. Therefore, if the particle undergoes many stochastic interactions, will finish gaining energy. That is to say, place a particle in a field of stochastic forces, any that this be, is a way to accelerate it. This seems paradoxical considering that since if in each interaction the particle can gain or to lose energy, one would be able to think that on the average does not gain neither lose energy and remains without change. Nevertheless the result of this study is that it does not matter which be the field of forces, if is random and isotropic, its effect will be an acceleration, a gain of energy.

## 2. Cones of gain and loss of energy

We know that if the force that acts on a particle aims in the same direction of the movement of the particle then the particle will gain energy. If the direction of the force is in contrary direction then loses energy. This carries us to the following questions: For a particle with a given initial moment $P_0$: what is the set of directions in which the force should act and the particle will gain energy when it receive a random impulse? And which is the set of directions in which the particle loses energy?

Let assume that X = {w|w(x,y,z)} is a region of the space, we also suppose a field of random force F(x,y,z,t) in the space X. The field of force F is random in magnitude and direction. The magnitude of F can have any distribution of probability but the distribution of directions is uniform, i.e., all the directions have the same probability, that is to say, any vector F can go in any direction with the same probability. Here we will not consider anisotropic fields that have not a distribution of uniform directions.

Let suppose that we have a particle with an initial moment $P_0$ at time $t_0$ in the point w(x,y,z) $\in$ X. On the particle acts the random field of force F that exert an thrust T in it. In the momentum space the interaction between the particle and the field is represented mathematically by the sum of $P_0$ and T and gives the $P_f$ moment after the interaction (see figure 1). Taking a given $P_0$ and T, the angle $\theta$ between the direction of $P_0$ and the direction of T has a fixed value $\theta_0$ for the case in which $|P_0|=|P_f|$. In this case the particle neither gains nor loses energy (figure 1C). When $\theta < \theta_0$ we have $|P_0|<|P_f|$, and then the particle gains energy (figure 1A).

There is another case in which $|T|>|2P_0|$. In this case the angle $\theta_0$ do not exists because in all the angles the particle will gains energy (figure 1D).

We can calculate the angle $\theta_0$ in the next way. If $P_0 = P_0\mathbf{i}$ and $T = T_x\mathbf{i} + T_y\mathbf{j} + T_z\mathbf{k}$ the final momentum ($P_f$) is:

$$\mathbf{P_f}=\mathbf{P_0}+\mathbf{T}=(P_0+T_x)\mathbf{i} + T_y\mathbf{j} + T_z\mathbf{k} \quad (1)$$

If $|P_0|=|P_f|$ then:

$$P_0^2 = (P_0+T_x)^2+T_y^2 + T_z^2 = P_0^2 + 2P_0T_x + T_x^2 + T_y^2 + T_z^2 \quad (2)$$

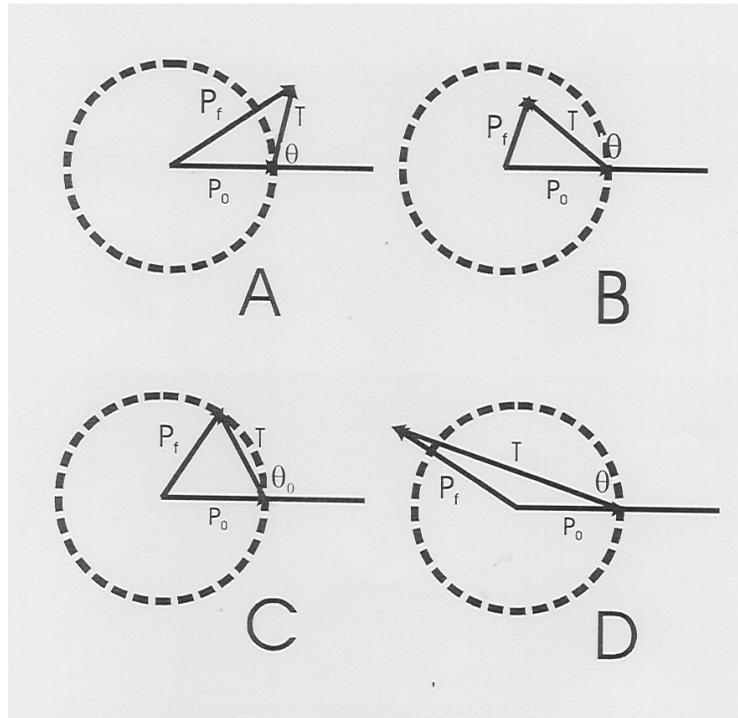

**Figure 1** The four diagrams show the sum, in the momentum space; of the initial momentum of the particle $P_0$ with the thrust T exerted by the field of forces. The broken circle is the cross section of the sphere with the radio equal to the magnitude of $P_0$. **(A)** If the arrow of the final momentum $P_f$ is outside of the discontinuous circle then $|P_0|<|P_f|$ and the particle gains energy. **(B)** If the arrow of the final momentum $P_f$ is inside of the discontinuous circle then $|P_0|>|P_f|$ and the particle loses energy. **(C)** If the arrow of the final momentum $P_f$ is on the discontinuous circle then $|P_0|=|P_f|$ and the energy of the particle does not change. If we have a $P_0$ and a given T, the angle $\theta$ between the direction of $P_0$ and the direction of T, when $|P_0|=|P_f|$, is fixed $\theta_0$. When the particle gains energy we have that $\theta < \theta_0$, and when the particle loses energy we have that $\theta > \theta_0$. **(D)** If $|T|>|2P_0|$ the arrow of the final momentum $P_f$ is outside of the broken circle in all the directions of T, and then the particle always gains energy.

When $\theta > \theta_0$ the relationship between $P_0$ and $P_f$ is $|P_0|>|P_f|$ and then the particle loses energy (figure 1B).

$$\Rightarrow \quad 0 = 2P_0 T_x + T^2 \qquad (3)$$

Where $T = T_x^2 + T_y^2 + T_z^2$ is the magnitude of **T**.

arXiv 18 April 2012

$$\Rightarrow \qquad \frac{T_x}{T} = -\frac{T}{2P_0} \qquad (4)$$

But $T_x/T = \cos\theta_0$ and then:

$$\cos(\theta_0) = -\frac{T}{2P_0} \qquad (5)$$

If T > $2P_0$ the function cosine is not defined and therefore do not exists a direction of deceleration; in all the directions of T the particle will gain energy (see figure 1D).

When **T** and **P₀** have a angle $\theta_0$. We can rotate **T** taking the direction of **P₀** as axis of rotation and **T** generates a conical surface that divides the unitary radio sphere of directions in two cones (see figure 2). ADECA is the acceleration cone; if T is in any direction inside this cone, the particle gains energy. The cone ABECA is the cone of deceleration; if **T** is in any direction, inside this cone, the particle loses energy. If T > $2P_0$ then does not exist cone of loss and the particle only can gain energy.

**3 Probability of gain and loss of energy**

But, how can we calculate the probability that a random thrust, can cause a particle gain or lose energy? The probability is defined like the reason of the quantity of favorable cases to the quantity of total cases, and therefore we can define the probability that the particle gain ($P_g$) or lose ($P_l$) energy in the following way:

$$P_g = \frac{\Psi_g}{\Psi} \qquad (6)$$

$$P_g = \frac{\Psi_l}{\Psi} \qquad (7)$$

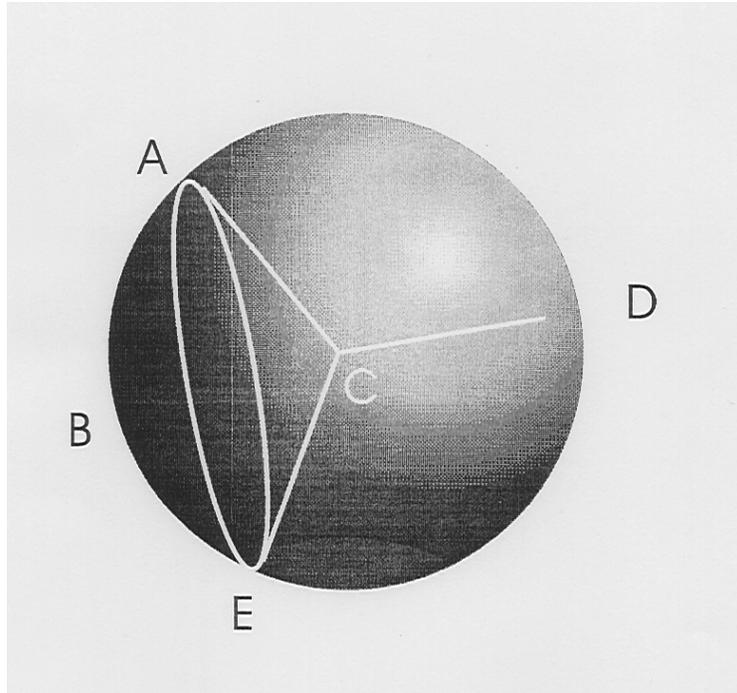

Figure 2. Unitary sphere of directions. C is the center of the sphere. CD is the direction of P0 and CA is the direction of T at angle ACD =$\theta_0$, the angle at which the force maintains the same energy of the particle. This angle divides the space of directions in two cones. ADECA is the cone in which the particle gains energy, and ABECA is the cone in which the particle loses energy. Of the fact that the cone of gain be larger than the cone of loss can be inferred that the gain of energy is going to prevail over the loss.

Where $\Psi_g$ is the solid angle of the cone of gain, $\Psi_l$ is the solid angle of the cone of loss, and $\Psi$ is the solid angle of the complete sphere of directions. If we take the unitary sphere of directions; the solid angle ($\Psi_g$) is equal to the surface of the sphere limited by the cone of gain, $\Psi_l$ is equal to the surface of the sphere limited by the cone of loss and $\Psi = 4\pi$ is all the surface of the sphere. From the formula of the area of a spherical segment (Bronshtein and Semendiaev, 1977) it is obtained:

$$\Psi_l = \pi(2+2\cos\theta_0) = 2\pi(1 + \cos\theta_0) \qquad (8)$$

Since $\Psi_g + \Psi_l = 4\pi$ then:

$$\Psi_g = 4\pi - \Psi_l = 4\pi - 2\pi(1 + \cos\theta_0) = 2\pi(1 - \cos\theta_0) \qquad (9)$$

And then from equations (6), (7), (8), y (9) we obtain:

$$P_g = \tfrac{1}{2}(1 - \cos\theta_0) \qquad (10)$$

$$P_l = \tfrac{1}{2}(1 + \cos\theta_0) \qquad (11)$$

And using the equation (5) we obtain the probability in function of the magnitude of the thrust (T) and the magnitude of the initial momentum ($P_0$) of the particle:

$$P_g = \tfrac{1}{2}(1 + T/2P_0) \qquad (12)$$

$$P_l = \tfrac{1}{2}(1 - T/2P_0) \qquad (13)$$

## 4. Results and Conclusions

In the figure 3 we can see the graphic of the equations (12) and (13) in function of the ratio $T/2P_0$.

When the reason $T/2P_0$ takes the value 0, the probability of gain ($P_g$) and the probability of loss ($P_l$) take the same value $\tfrac{1}{2}$. Since the probabilities are equals this signifies that, in a large number of interactions of the particle with the field, the gained energy is going to be equal to the loss energy and therefore the particle, on the average, conserves its initial momentum. The quotient $T/2P_0$ takes the value 0 in two cases; the trivial case when T = 0 and therefore the force does not exist and the particle maintains its energy. The second case is when T ≠ 0 and $P_0 \gg T$; strictly speaking, in the limit as $P_0$ approaches to infinite. Of this result we can conclude that for an interval of given values of T, if $P_0$ is a larger that T, that is to say for large velocities of the particle, the value of $P_g$ is nearby al value of $P_l$, and both close to $\tfrac{1}{2}$ and therefore the gain of energy is similar to the loss and, at long-term, the particle does not gains neither loses energy.

But for $T/2P_0 > 0$ we have an important result that is that the probability of gain ($P_g$) is larger that the probability of loss ($P_l$) for all the values of the ratio. Of this result we can conclude that for intermediate values ($0 < T/2P_0 < 1$) of the ratio, if the particle receives many random accelerations, at long-term it will gain energy, independently that in each interaction it can gain or lose energy.

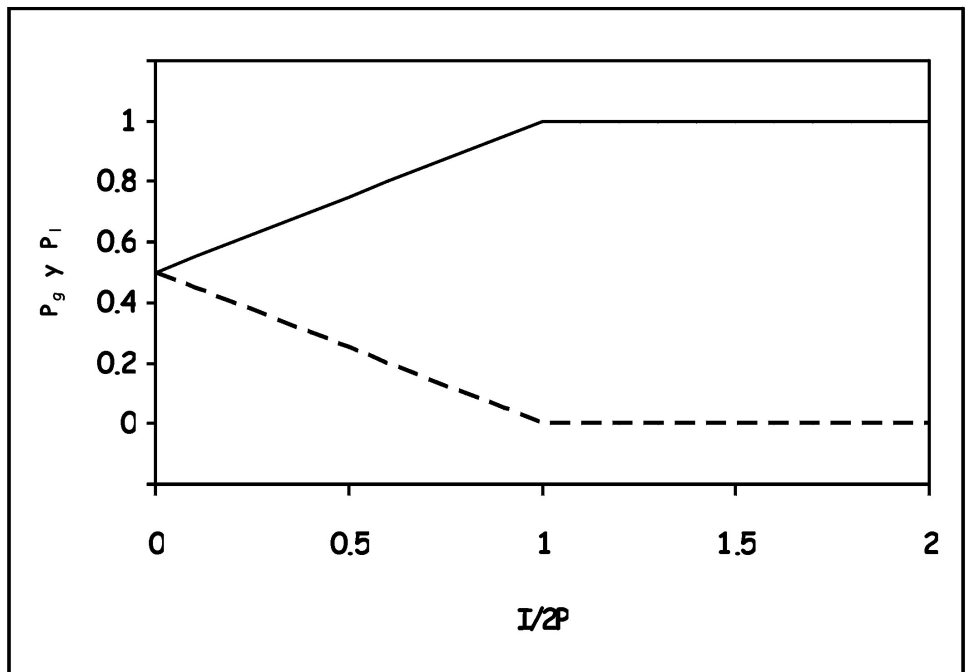

Figure 3. Probability vs. $T/2P_0$. Graphic of the equations (12) and (13). $P_l$ is the dashed line and $P_g$ is the continuous line. For all the values of $T/2P_0$ except 0, in which $P_l = P_g = 0.5$, the two probabilities have the following relation: $P_l < P_g$.

For values > 1 of this ratio, the cone of deceleration do not exists and therefore $P_g = 1$ and $P_l = 0$, that is to say, for low velocities the particle gains energy in all the interactions.

An important implication of the prior results is that these results are independent of the type of distribution that has the intensity of the random field of force.

arXiv 18 April 2012

Of the previous considerations we can induce that the stochastic acceleration is a mechanism that has its maximum of efficiency when the particle has velocity zero, and therefore $P_g = 1$, and goes losing in efficiency when the particle increases its velocity and $P_g$ approaches the value ½.

These results are strongly dependent of the uniform distribution of directions. There are not valid for a different distribution of directions, since when there are anisotropies in the directions field can be presented tendencies that clearly cause preferential losses or gains of energy. For example a field of forces whose directions only fluctuated between θ0 and $\pi$ clearly would decelerate the particles.